# All-electrical valley filtering in graphene systems (II): Numerical study of electron transport in valley valves

Jia-Huei Jiang,[1] Ning-Yuan Lue,[2] Feng-Wu Chen,[1] and Yu-Shu G. Wu[1,2,†]

[1] *Department of Physics, National Tsing-Hua University, Hsin-Chu 30013, Taiwan, ROC*
[2] *Department of Electrical Engineering, National Tsing-Hua University, Hsin-Chu 30013, Taiwan, ROC*

This work performs a numerical study of electron transport through the fundamental logic gate in valleytronics – a valley valve consisting of two or increasing number of valley filters. Various typical effects on the transport are investigated, such as those due to interface scattering, long- and short- range impurity scattering, edge roughness, strain, inter-filter spacing, or increasing number of valley filters. For illustration, we consider the class of specific valves built from graphene quantum wire valley filters in single layer or bilayer graphene, with the filters subject to separate control of in-plane, transverse electric fields. The nearest-neighbor tight-binding model of graphene is used to formulate the corresponding transport problem, and the algorithm of recursive Green's function method is applied to solve for the corresponding transmission coefficient. In the case of two-filter valves, the result explicitly demonstrates the existence of a pronounced on-off contrast in electron transmission between the two configurations of valves, namely, one with identical and the other with opposite valley polarities in the two constituent filters. The contrast is shown to be enhanced when increasing the number of filters in valves. Signatures of Fano-Fabry-Perot type resonances in association with interface scattering and inter-filter spacing are illustrated. Electron backscattering due to impurities is found to be sizably suppressed, with the valve performance showing considerable robustness against edge roughness scattering. On the other hand, the presence of a uniaxial strain modifies the electron transmission and results in an interesting quasi-periodic modulation of transmission as we vary the strain strength.

[†] Corresponding author. Email: yswu@ee.nthu.edu.tw

## I. INTRODUCTION

Graphene electrons close to the Dirac points carry a novel degree of freedom - valley pseudospin with rich electrical, magnetic and optical properties [1–4]. Such an inherent degree of freedom emerges in 2D materials with hexagonal symmetry, due to the existence of two degenerate and inequivalent band structure valleys (K and K') that transform into each other under time-reversal symmetry operation. As such, a graphene electron is characterized by a binary valley pseudospin index, e.g., $\tau_v$ with $\tau_v = +1 / -1$ for K / K' valley. This index is analogous to the electron spin index $\sigma_z$ for up / down state, with the difference that while the quantization axis for $\sigma_z$ can be arbitrarily chosen, there is no such freedom for $\tau_v$.

The existence of valley pseudospin opens up a new type of electronics - valleytronics which can be built to utilize the valley pseudospin either as a binary information carrier or for logic gate implementation. With the corresponding device principle different from those of charge- or spin- based devices, valleytronics expands the electronics family and introduces additional flexibility to electronic device design. Concrete proposals of graphene-based valleytronic devices have emerged such as those of valley valves [1,5,6], which can turn on or off electrical currents like field effect transistors (FETs), and therefore can serve as the fundamental block for building logic gates and circuits. Other examples of fundamental valleytronic devices include valley qubits [7,8] and valley FETs [9]. Moreover, valleytronic materials beyond graphene have also been explored, such as transition metal dichalcogenides which exhibits a strong spin-orbit interaction and spin-valley coupling further enriching the field of valleytronics [10–12].

From the perspective of device downscaling, valley valves are of great interest, since the on-off switching principle of such valves differs from that of conventional FETs and could significantly reduce power consumption – an essential requirement for device downscaling. A valley valve consists of valley filters each of which separately generates a valley-polarized electrical current. When two such filters are combined in series, they form a valve that passes or blocks the current depending on the relative polarity between valley currents generated by the filters. Specifically, the configuration of identical valley polarities passes a high current and results in the "on" state and that of opposite polarities blocks the current and results in the "off" state. The on-off contrast in current magnitude depends on many factors and constitutes an important figure of merit for a valley valve.

The present work belongs to a series of our recent studies in ballistic valley filters and valves, which begin with a proposal of all-electrical valley filters [13] (referred to as **W-I** below), and continue with the present work (referred to as **W-II** below) to demonstrate various practical effects in generic valley valves and in the valves considered in **W-I**.

A number of valley filters or valves have been proposed or investigated. Among them, some are based on the chiral mode of zigzag graphene nanoribbons [1], some on the valley Hall effect [2,3,6,14–16], some on the electron scattering off a line defect [17–19], some require the presence of strain or magnetic field [20–24], some achieve valley polarization with intense THz light [25], and some make use of the band structure warping [26,27]. For the demonstration of various effects in valley valves, we consider the class of valves formed of the ballistic filters proposed in **W-I** with the following desirable features, namely, i) all-electrical controllability, ii) time-reversal symmetry, and iii) valley-propagation correspondence. The first feature complies with the prevailing practice in present semiconductor industry. The third feature is connected with the second, as discussed in the following. Let $\tau_v$ be the primary valley character carried by an electron moving with wave vector **k**. Then a time-reversal operation on the electron would flip **k** to –**k** and at the same time $\tau_v$ to –$\tau_v$, which is the manifestation of valley-propagation correspondence. As shown in **W-I**, such filters can be built with graphene quantum wires that utilize the mechanism of valley-orbit interaction (VOI) for valley filtering. As will become clear below, a large part of the effects studied in this work with VOI based valves are expected to be generic in valves characterized by the three aforementioned features. Examples of such effects include those due to impurity scattering, interface scattering, inter-filter spacing, and increasing number of filters in valves.

A VOI-based valley valve features i) a quasi-one diemsional (Q1D) electron transport channel in the armchair direction in gapped graphene, ii) defect scattering near the channel, and iii) in-plane electric fields transverse to the channel for separately controlling electron transport in constituent filters. From the fabrication perspective, a VOI-based valve can be realized in single layer graphene (SLG). For example, the Q1D channel can be created by placing SLG on a trenched h-BN substrate. In such a structure the suspended part of SLG above the trench is gapless and, electronic motion in this part is confined on both sides by h-BN substrate-supported, gapped SLG [3,28,29], thus giving rise to the formation of a Q1D channel. Alternatively, one could use the BN-doped SLG $(BN)_xC_{1-x}$. With the energy gap being composition-dependent [30], one could achieve a gap profile to effect the required quantum confinement with, for example, a lateral structure such as $(BN)_xC_{1-x}/(BN)_yC_{1-y}/(BN)_xC_{1-x}$ $(BN)_xC_{1-x}$. On the other hand, such a valve can also be realized in AB stacked bilayer graphene (BLG). **Figure 1** shows a prototype VOI-based valve in AB stacked BLG. Each constituent filter consists of a graphene quantum wire in the armchair direction with full electrical maneuverability. Each filter is delimited on two sides by lines of scattering defects in the x-direction. Such defect lines can be realized by impurity implantation, graphene oxidation, or simply edges of an arm-chair graphene nanoribbon (AGNR). As shown in the figure, each filter is a gated structure. The vertical bias difference between top and bottom gates opens up the energy gap of BLG [31] below top gates in such a way that the energy gap there is made larger than that in the channel. Therefore, graphene there plays the role of potential barrier providing quantum confinement in the y-direction to carriers in the channel. Moreover, a bias difference between the pair of split top gates can generate an in-plane electric field ($\varepsilon_y$) in the y-direction.

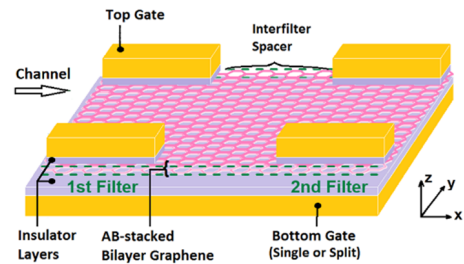

**Figure 1.** The basic construct of a VOI valley valve in AB stacked bilayer graphene with layers lying in the x-y plane. In each filter the electron transport is conducted in a quasi-1D channel in the x-direction defined by top and and bottom gates. Dashed lines denote defects.

Next, we briefly describe the principle underlying VOI-based filters. Valley filtering in such filters is controlled by $\varepsilon_y$, as briefly explained below by using a symmetry argument and the perturbation theory. First, we ignore the effect of defect lines and consider a channel electron state characterized by ($k_x$, $\tau_v$). In the absence of $\varepsilon_y$, the structure possesses the symmetry of reflection with respect to the channel's horizontal symmetry axis. This results in degenerate energy sub-bands, e.g., $E(k_x; \tau_v = +1) = E(k_x; \tau_v = -1)$. When $\varepsilon_y$ is present, the symmetry is broken leading to $E(k_x; \tau_v = +1) \neq E(k_x; \tau_v = -1)$, the so-called "Rashba valley splitting" as depicted in **Figure 2(a)**, where the bands are split but cross each other at $k_x = 0$. Note that the crossing at $k_x = 0$, i.e., $E(k_x = 0; \tau_v = +1) = E(k_x = 0; \tau_v = -1)$ follows from the time-reversal symmetry. Next, we take into account the effect of defect scattering near the channel. The presence of the scattering results in an inter-valley coupling between K and K' electron states, which mixes the states as well as modifies the dispersion in such a way that the crossing energy subbands of opposite valleys now become anti-crossing, with a corresponding pseudogap opened between the band edges at $k_x = 0$ of the two subbands as shown in **Figure 2(b)**. Specifically, the induced valley mixing is strongest at $k_x = 0$. Since $E(k_x = 0; \tau_v = +1) = E(k_x = 0; \tau_v = -1)$ without the mixing, the mixing leads to 50:50 valley mixed states that are symmetric or antisymmetric combinations of the two valleys, according to the degenerate perturbation theory. However, away from the point, since $E(k_x; \tau_v = +1) \neq E(k_x; \tau_v = -1)$, the mixing is limited, according to the nondegenerate perturbation theory, with the resultant mixed state being dominated by the original valley character of the state before mixing. Overall, it achieves valley filtering with the following feature, namely, that the pseudogap in the modified dispersion forms an energy window of valley polarized electron states, where right-moving electrons are primarily polarized into one valley while left-moving ones into the other.

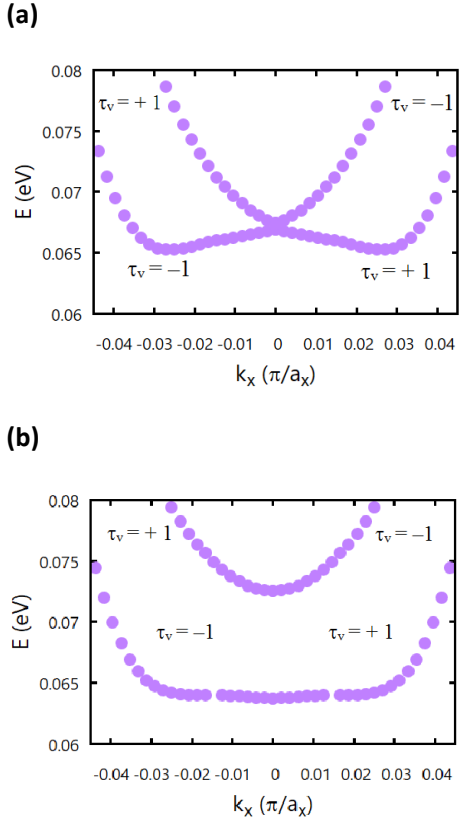

**Figure 2.** (a) Rashba valley splitting. (b) Formation of a pseudogap. The structure considered here is a quantum wire channel embedded in a single layer AGNR, with both the wire and AGNR aligned along the x-direction. "$a_x$" is the lattice constant. Structural parameters are channel width = 5.04 nm; channel gap = 0.1eV; barrier gap = 0.4 eV; conduction band offset = 0.15 eV; potential energy drop across the channel = 0.1 eV; and barrier width = ∞ (9.84 nm) in (a) ((b)).

In fact, one can go beyond the symmetry argument and derive the functional form of valley splitting. For example, in the case of SLG, the derivation shows the emergence of a VOI term [7,32,33] in the electron Hamiltonian given by

$$H_{vo} = \frac{\hbar}{4m^*\Delta}(\nabla V \times \vec{p}) \cdot \tau_v \hat{z}$$

($2\Delta$ = energy gap, V = electric potential energy, $\vec{p}$ = electron momentum operator, and $\hat{z}$ =

out-of-plane unit vector). With $V = -e\varepsilon_y y$, it leads to the simple expression $H_{Rashba} = \alpha \tau_v k_x \varepsilon_y$ ($\alpha$ = Rashba constant) for the splitting. A similar but more complicated expression also exists in the case of bilayer graphene [34]. Note that $H_{Rashba} \propto \tau_v \varepsilon_y$, meaning that valley characters of the split bands are actually determined by the field direction – a reversal of the direction switches valley characters of split bands. Correspondingly, it means that the valley polarity of a VOI-based filter is electrically switchable.

VOI-based valley filters can be utilized to realize valley valves. As shown in **Figure 1**, by placing two pairs of split top gates outside the Q1D channel, two filters are effectively combined in series forming a valve. The electron transmission through it can be controlled by applying biases to the two pairs of gates. High (low) transmission is obtained if the biases create parallel (antiparallel) in-plane fields. Throughout this work we shall use P (AP) to denote the configuration of parallel (antiparallel) fields.

This work performs the numerical study of electron transmission through a valley valve, with the VOI-based valve being the representative system. Generic effects are studied such as those due to interface scattering, long- and short- range impurity scattering, inter-filter spacing, or increasing number of filters in valves. In addition, we also look into effects specific to VOI-based valves such as those due to strain or edge roughness. We formulate the corresponding transport problem within the nearest-neighbor tight-binding model of graphene lattice, and employ the algorithm of recursive Green's function method to calculate the transmission. The result explicitly demonstrates the existence of a large on-off contrast in electron transmission between the P and AP configurations of two-filter valves. The contrast is shown to be enhanced in three-filter valves. Moreover, signatures of Fano-Fabry-Perot type resonances induced by multiple interface scattering are illustrated, in valves with inter-filter spacing or more than two filters. In general, due to the valley-propagation correspondence in constituent filters, electron backscattering is found to be sizably suppressed leading to robustness in the transmission against limited impurity or edge roughness scattering. On the other hand, the presence of a uniaxial strain modifies the electron transmission and results in an interesting quasi-periodic modulation of transmission with varied strain.

Two points in the study are noted below. Firstly, simulations throughout the work are all performed at zero temperature for simplicity. However, downgraded performance is expected when raising the temperature of the structure, as electrons start to occupy states above the pseudogap giving a current of valley polarization opposing that due to states below the gap. While increasing the pseudogap size may offer a feasible solution to the issue, the approach may have to pay, as noted in **W-I**, for the cost of reduced state valley polarization. Therefore, overall, a trade-off generally exists between the performance and operational temperature. Secondly, the study is focused on relatively narrow channels. Admittedly, if we use wider structures, the number of subbands involved would increase with varied valley polarizations leading to downgraded performance. However, since an important mission of valleytronics is to resolve issues facing traditional device downscaling, it justifies the focus on relatively narrow channels.

This article is organized as follows. In **Sec. II**, we describe the various structures considered in the study. In **Sec. III**, we describe the theoretical method. In **Sec. IV**, we present the numerical results. In **Sec. V**, we summarize the study.

## II. STRUCTURES

In this section, we describe the structures considered in **Sec. IV** for the discussion of various effects in valves.

### 1. General considerations

**Figure 3** presents the top view of basic layout of a VOI valley valve. Various parameters are shown in the figure and the same notations will be used throughout this work. Although defect lines on top and bottom sides can be realized in different ways, in our study they are taken to be nanoribbon edges. Unless specified otherwise, the valve is patterned in an AGNR in the 3M family (M = number of atomic rows). Such nanoribbons are known to be semiconducting even if the corresponding bulk gap vanishes [35]. We take the bulk gap profile transverse to the channel to vary in such a way that the channel is confined on both sides by wide gapped barriers.

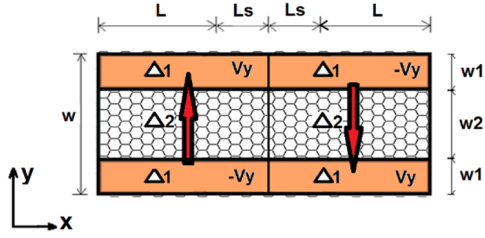

**Figure 3.** Top view of a VOI valley valve with back-to-back valley filters. Dashes lines on top and bottom sides denote defect lines. w = total width, w1 = barrier width, w2 = channel width, L: buffer region, and $L_s$: reserved range for impurity or edge roughness distribution. $2\Delta_1$ = barrier graphene gap, $2\Delta_2$ = channel graphene gap, and $2V_y$ = transverse potential energy difference. Vertical arrows denote electric fields in the AP configuration.

In applications, a valve is connected to electrodes on both ends. In all calculations we disregard electrode effects and focus on the valley-dependent transport in valves. To achieve this, we take the filters adjacent to electrodes to be nearly semi-infinite or, equivalently, place valve/electrode interfaces at remote distances and thus ignore secondary effects in association with valve/electrode interfaces on valley transport.

### 2. Single layer- and bilayer- graphene based structures

Valves in both SLG and BLG are studied. With the existence of VOI in gapped SLG or BLG as mentioned earlier, graphene in either form can be used for the implementation of VOI-based valleytronics. However, from the computational point of view, since a SLG structure contains only half as many atoms as the corresponding BLG structure, there is a difference - a SLG structure would greatly facilitate the calculation and illustration of valley transport. Therefore, part of our study is performed with SLG-based valves as effective models with adjustable parameters, e.g., band gaps and offsets, when it focuses on universal effects insensitive to the graphene layer number. On the other hand, due to the presence of interlayer coupling in BLG, SLG and BLG do exhibit different, layer number-dependent properties. For example, in the case of bulk band structures, SLG shows near each Dirac point the monotonous pseudo-relativistic energy dispersion described by $E_\pm(\vec{k}) = \pm(\Delta^2 + \hbar^2 v_F^2 |\vec{k}|^2)^{1/2}$, with "+" for the conduction band and "-" for the valence band, while BLG shows a non-monotonous energy dispersion with a Mexican hat structure near each Dirac point. As the electron transport depends on underlying bulk band structure, details of the transport are expected to vary between SLG- and BLG-based valves.

### 3. Type-I and Type-II channels

**Figure 4** shows both conduction band edge (CBE) and valence band edge (VBE) profiles transverse to the channel. According to these profiles, VOI valves are classified into those

**(a)**

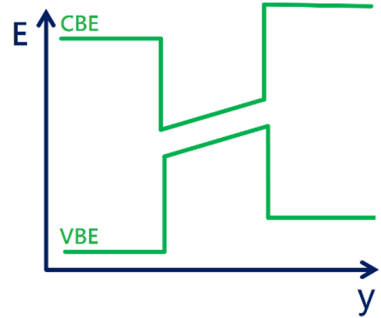

**(b)**

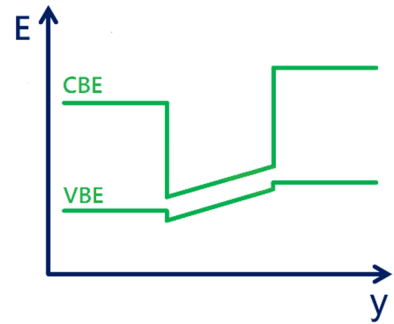

**Figure 4.** Transverse profiles of conduction band edges (CBE) and valence band edges (VBE) for **(a)** type-I channel and **(b)** type-II channel.

with Type-I channels when straddling alignment occurs and those with Type-II channels when staggered alignment occurs. Generally speaking, valley-dependent transport in lateral quantum structures of 2D materials depends on the band edge alignment [13,36]. Therefore, electron transmissions through valves with these two types of channels are investigated in this work.

Four structures, SLG-I, BLG-I, SLG-II, and BLG-II will be frequently used in this work. SLG-I denotes the valve with a Type-I channel in SLG specified by the following parameters: barrier width = 9.84 nm; channel width = 5.04 nm; barrier gap = 0.4 eV; channel gap = 0.1 eV;

conduction band offset = 0.15 eV; transverse potential energy drop = 0.1 eV; L (length of buffer) = 426 nm; and $L_S$ (length of reserved range for impurity or edge roughness distribution) = 170 nm. BLG-I denotes the valve with a Type-I channel in BLG specified by the following parameters: barrier width = 4.92 nm; channel width = 7.5 nm; barrier gap = 0.32 eV; channel gap = 0.1 eV; conduction band offset = 0.11eV; transverse potential energy drop = 0.06 eV; L = 426 nm; and $L_S$ = 170 nm. SLG-II denotes the valve with a Type-II channel in SLG specified by the following parameters: barrier width = 8.61 nm; channel width = 7.5 nm; barrier gap = 0.21 eV; channel gap = 0.2 eV; conduction band offset = 0.12 eV; transverse potential energy drop = 0.1 eV; L = 426 nm; and $L_S$ = 170 nm. BLG-II denotes the valve with a Type-II channel in BLG specified by the following parameters: barrier width = 5.16 nm; channel width = 5.16 nm; barrier gap = 0.2 eV; channel gap = 0.18 eV; conduction band offset = 0.12eV; transverse potential energy drop = 0.08 eV; L = 426 nm; and $L_S$ = 170 nm. Unless specified otherwise, this work uses the same parametric sets as those given above. We note two points. First, in BLG-I and BLG-II, band gap parameters are chosen to reflect the values presently accessible in experiments [37]. Band offsets are chosen in such a way that they are achievable with suitable design of gated structures as discussed in **W-I**. Transverse potential energy drops are chosen to attain large VOI strengths while subject to the constraint of preserving quantum confinement of channel carriers. Second, SLG-I and SLG-II can also serve as effective models to allow for insights into universal features of valley-dependent transport in valves.

4. **Inter-filter spacers**

In a practical valve, the two constituent filters are actually placed slightly apart from each other, in order to avoid cross-interference. As a result, there is an inter-filter spacer between the two constituent filters, as shown in **Figure 5**. We denote the inter-filter spacing as $L_{IFS}$. For short $L_{IFS}$, the spacer is expected to serve as a passive channel for electron transport, and its presence would only slightly alter the transmission. However, for long $L_{IFS}$, nontrivial effects on the transmission are expected due to multiple interface scattering at the two filter/spacer interfaces, which can result in Fano-Fabry-Perot type resonances modifying the transmission. In order to study the effects, we shall vary the spacer length. Since Fano-Fabry-Perot resonances are tied to resonant modes in the spacer, varying the length would modulate such modes and allow us to probe the effects. In addition, we shall also study the case where the spacer is specifically formed of an armchair nanoribbon. In this case, when an electron passes through the spacer, additional edge scattering would be incurred that causes inter-valley coupling [38]. As a result, the valley current is expected to be depolarized when passing through the spacer. We shall examine such an effect in **Sec. IV** with a contrast study of the three cases shown in **Figure 5**, where the edge scattering effect is tuned by placing the edges at different distances from the channel's horizontal symmetry axis. As shown in **Figure 5**, the edges are pushed outward when going from **Figure 5(a)** to **Figure 5(b)**, and blocked by barriers from passing electrons in **Figure 5(c)**.

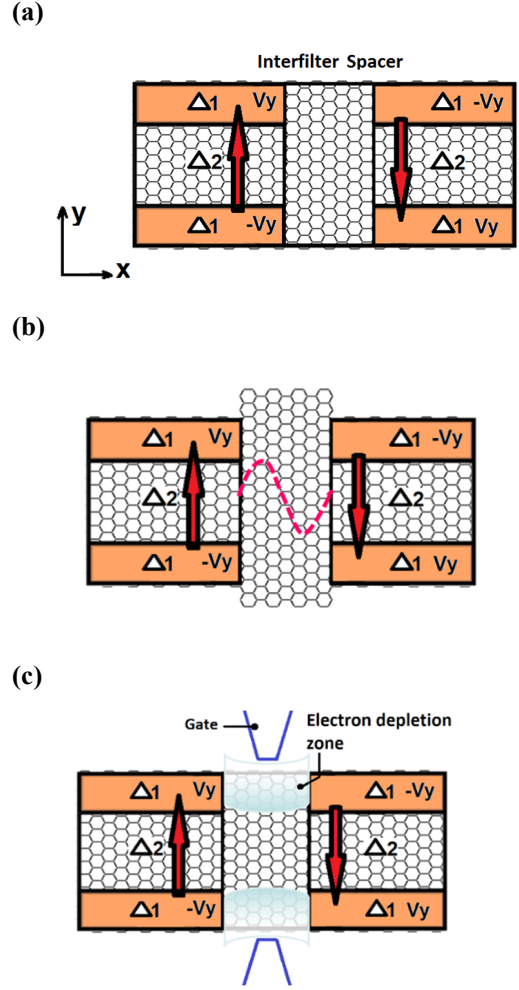

**Figure 5.** Valley valves with inter-filter spacers. **(a)** The spacer is a nanoribbon as wide as the filters. **(b)** The spacer is a nanoribbon wider than the filters. A resonant mode in the spacer is illustrated in **(b)**, with the red dashed line being the corresponding amplitude profile in the x-direction. **(c)** The spacer is a Q1D channel formed with gate-induced confinement barriers. Vertical arrows denote electric fields in the AP configuration.

We note that, except in the study of inter-filter spacer effect, this work takes the short spacing limit ($L_{IFS} = 0$) where the transmitted wave exiting a filter immediately enters the subsequent filter. Such an approach disregards the spacer-induced effect on electron transmission and isolates other effects of interest for study. By doing so, it eases the analysis of other effects.

## 5. Impurity scattering

Impurity scattering in graphene changes electron momentum and disturbs electron transport, which can lead to various interesting phenomena [39,40]. In the context of valley transport, impurity scattering calls for special attention, since in addition to intra-valley scattering, it induces inter-valley transfer that flips valleys and affects the transport.

Owing to the large wave vector difference between opposite valleys, however, impurity-induced inter-valley transfer is usually restricted. Such a restriction may be exploited to provide protection for electron transport against impurity scattering. In the case of a VOI-based valve, for example, the protection takes place because of the valley-propagation correspondence in such a valve. In analogy to the well-known spin-wave vector locking in topological insulators [41], electron valleys and wave vectors in the valve are locked, although imperfectly due to incomplete valley filtering in constituent filters. As a result, backscattering of electrons would require simultaneous valley and wave vector flipping. This constraint suppresses the scattering in favor of robust electron transmission.

The valley-wave vector locking-induced protection of electron transmission is investigated as follows. Let $V_{imp}(r)$ = impurity potential. Then, the strength of valley flipping scattering is roughly determined by the matrix element $V_{imp}(K-K')$, the Fourier transform of $V_{imp}(r)$ that strongly depends on the range of $V(r)$. We thus conduct the investigation for both long- and short- range impurities. We take impurity potentials to be given by the following Gaussian function [39,42]

$$V_{imp}(\vec{r}) = \sum_i u_i \exp(-\frac{|\vec{r}-\vec{r}_i|^2}{d^2}) \quad (1)$$

where $\vec{r}$ is the position vector of an electron, $\vec{r}_i$ denotes the center of a randomly distributed impurity, $u_i$ is the corresponding scattering strength randomly distributed in the interval (-$u_M$, $u_M$), and $d = 1.5\sqrt{3}a_0$ ($d = 0.1\sqrt{3}a_0$) for long-range (short-range) scatterers ($a_0$ = the carbon-carbon distance). $u_M$ is determined by the following normalization condition:

$$u_M \sum_{\vec{r},\vec{r}'} \exp(-\frac{|\vec{r}-\vec{r}'|^2}{d^2}) = u_0 \quad (2)$$

where we choose $u_0 = 0.5$ eV.

In order to study the impurity effect, we place impurities only in the first filter but leave the second filter undoped, with the second filter serving to detect the impurity-scattered valley polarized current exiting the first, as shown in **Figure 6**. This gives us a way to probe the impurity scattering effect in the first filter and thus also in the valve.

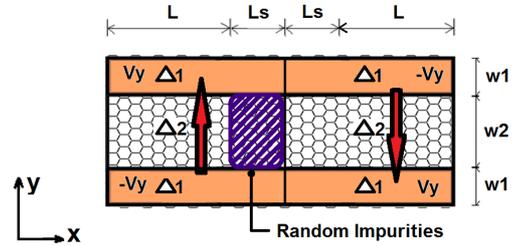

**Figure 6.** An illustrative valve configuration for the study of impurity scattering. Vertical arrows denote electric fields in the AP configuration.

## 6. Edge roughness

Apart from impurity scattering, another frequently encountered issue in nanoribbon systems is edge roughness. We investigate the effect of such scattering in nanoribbon-based valley valves, within the model illustrated in **Figure 7** with a single layer AGNR. As shown in **Figure 7(a)**, edge roughness is included only in the first filter. The second filter serves to detect the edge roughness-scattered valley polarized current exiting the first. As illustrated in **Figure**

**7(b)**, the edge imperfection considered in this work appears in the form of carbon vacancy clusters.

Edge roughness model parameters here are (i) the probability of an edge vacancy cluster occurring ($p_{vc}$), and (ii) the width of a vacancy cluster ($\Delta W$). $\Delta W$ is taken to be either one or two atomic rows. For these two kinds of clusters, the probability of occurrence is taken to be the same, thus introducing a random nature to the vacancy cluster distribution. It follows that the mean width $\overline{\Delta W}$ is given by

$$\overline{\Delta W} = \frac{\sqrt{3}}{2} a_0 \frac{1+2}{2} \approx 0.184 \; nm.$$

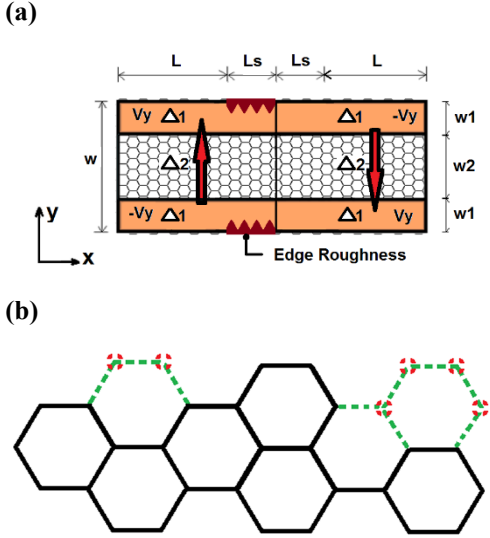

**Figure 7.** Edge roughness. **(a)** An illustrative valve configuration for the study. Vertical arrows denote electric fields in the AP configuration. **(b)** Types of vacancy clusters on ribbon edges, with the illustration showing one row and two rows wide clusters. Dashed green lines: broken bonds; dashed pink circles: carbon atom vacancies.

Generally speaking, edge roughness plays a similar role to impurities, in the sense that its main effect on valley transport comes from the electron scattering it induces, including that which couples opposite valleys. Interesting suppression of edge roughness scattering is therefore expected due to the valley-wave vector locking as in the case of impurity scattering. Details will be discussed in **Sec. IV**.

## 7. Three-filter valves

The idea of combining two valves in series is motivated by the practical need to enhance the on-off contrast in electron transmission. In this work, we investigate a three-filter valve formed from such a combination, as shown in **Figure 8**, where the first valve consists of the first and second filters and the second valve consists of the second and third filters. Note, for simplicity, that the three filters are connected back to back to exclude the inter-filter spacer effect from consideration.

In the ballistic regime, electron transport in the three-filter configuration is more than a sequential transmission through two independent valves. With the presence of filter/filter interfaces, resonant modes could appear inside the middle filter due to multiple interface scattering. Therefore, interference effects such as Fano-Fabry-Perot resonances could show up in company with transmission via such resonant modes.

In the study we shall take transverse field orientations in the first and third filters to be the same, and vary only the field orientation of middle filter. Such an arrangement is expected to pass a high current when the valve is configured with all three transverse fields aligned in the same direction (which is referred to as the P configuration of the three-filter valve below). On the other hand, in the configuration where the middle field is reversed (which is referred to as the AP configuration of the three-filter valve below), both the first and second valves are set at "off" state. As such, in this case it is expected to pass a low current. The performance of three-filter valves will be assessed in **Sec. IV** based on the corresponding on-off contrast between the two forgoing configurations of field orientations.

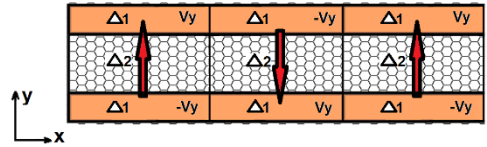

**Figure 8.** The layout of a three-filter valve with electric fields oriented for passing a low current.

## 8. Strained structures

Strain in graphene is known to give rise to interesting effects [35]. With armchair Q1D channels being an important part of VOI-based valley valves, we specifically investigate effects of uniaxial strain with corresponding axis being along or transverse to the channel. The study is performed within the nearest-neighbor hopping tight-binding model including the presence of uniaxial strain, where hopping from one atomic site to its three nearest neighbors no longer possesses three-fold rotational symmetry. Instead it requires the description with two hopping parameters, namely, $t$ for the hopping along armchair direction and $t'$ for that along zigzag direction, with the ratio $(t'-t)/t$ reflecting the magnitude of strain. In **Sec. IV**, the electron transmissions through a valve will be studied as a function of $t'/t$.

### III. THEORETICAL METHOD

In this section, we describe the theoretical method for the numerical study of VOI valley valves - the formulation of electron transmission problem in the tight-binding model. We also discuss the corresponding probability current operator for the computation of electron transmission.

Note that the formulation described is analogous to the Landauer-Buttiker/Fisher-Lee approach that has been developed and applied in pioneering studies of electrical conductance in mesoscopic structures [43–48]. As the formulation and associated algorithm applied in every work actually vary, evolve, and depend on the crystal structure, we present below the ones used in this study, specifically with the 2D hexagonal crystal of graphene in the focus.

In brief, we divide the structure into many unit cell slices, and express the transmission in terms of the Green's function propagator "$G_{N0}$" which connects the far incident-side cell 0 to the far outgoing-side cell N. In particular, $G_{N0}$ is obtained in this work by $N$ steps of iteration using the recursive Green's function algorithm. We sketch the formulation and algorithm below.

#### 1. Tight-binding model

As is well known, the spin-orbit interaction in graphene is extremely weak [35]. In addition, we assume that both magnetic fields and magnetic impurities are absent from the system. Therefore, the electron spin degree of freedom is taken to be frozen and removed from the consideration of electron transmission through the valve. For single layer AGNR, the spinless tight-binding Hamiltonian with nearest-neighbor hopping is given by [35]

$$H_{SLAGNR} = \sum_{m,i} h_{m,i} c^+_{m,i} c_{m,i}$$
$$-\gamma_0 \sum_{m,\langle i,j \rangle} [c^+_{m,i} c_{m,j} + c^+_{m-1,i} c_{m,j} + c^+_{m+1,i} c_{m,j}], \quad (3)$$

where $c_{m,i}$ annihilates an electron in the $2p_z$ orbital at site $i$ in unit cell $m$; $h_{m,i}$ is the on-site energy; $\gamma_0 = 3.0$ eV is the nearest-neighbor hopping; and $<i,j>$ runs over all nearest neighbor pairs within or between cells (See **Figure 9(a)**). For AB-stacked bilayer AGNR, the corresponding Hamiltonian is given by [31]

$$H_{BLAGNR} = \sum_{L=\{1,2\},m,i} h_{L,m,i} (a^{+(L)}_{m,i} a^{(L)}_{m,i} + b^{+(L)}_{m,i} b^{(L)}_{m,i})$$
$$-\gamma_0 \sum_{L=\{1,2\},\langle m,i;m',j \rangle_0} (a^{+(L)}_{m,i} b^{(L)}_{m',j} + h.c.)$$
$$-\gamma_1 \sum_{\langle m,i;m',j \rangle_1} (a^{+(1)}_{m,i} a^{(2)}_{m',j} + h.c.)$$
$$-\gamma_3 \sum_{\langle m,i;m',j \rangle_3} (b^{+(1)}_{m,i} b^{(2)}_{m',j} + h.c.) \quad (4)$$

where $a_{m,i}^{(L)}$ ($b_{m,i}^{(L)}$) annihilates an electron in the $2p_z$ orbital of a carbon atom on A (B) site indexed by $i$ in unit cell $m$ of layer $L$; $h_{L,m,i}$ is

(a)

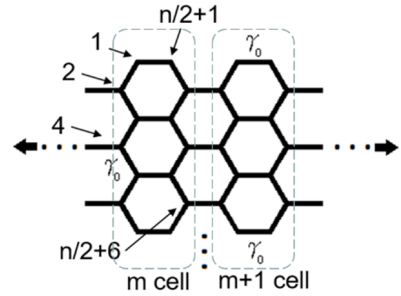

(b)

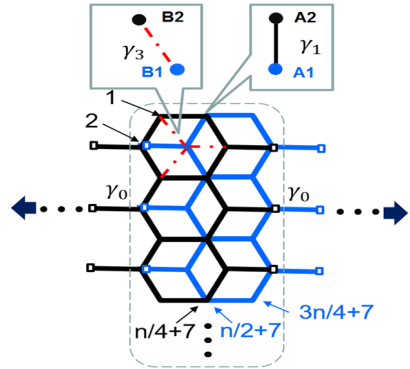

(c)

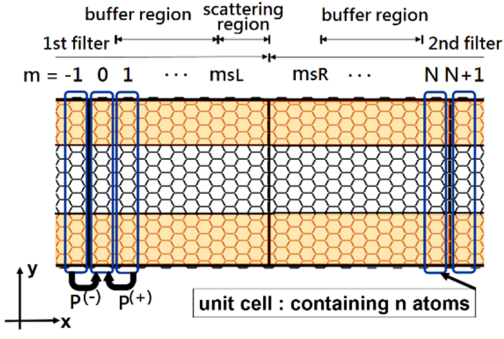

**Figure 9. (a)** Tight-binding description of single layer AGNR. **(b)** Tight-binding description of AB stacked bilayer AGNR. Dotted lines enclose respective unit cells. Atom indexing schemes are shown in **(a)** and **(b)** (with $n$ = total number of atoms in each cell). **(c)** Illustration of electron transmission problem for a valley valve, formulated in the tight-binding model. $N$ = total number of unit cells in the valve. $m$ = unit cell index. $m_{sL}$ = starting unit cell index of the region for impurity or edge roughness distribution. $P^{(+)}$ ($P^{(-)}$) denotes the nearest-neighbor coupling between $m^{th}$ and $(m+1)^{th}$ ($(m-1)^{th}$) unit cells.

the on-site energy; $\gamma_0 = 3.0$ eV, $\gamma_1 = 0.4$ eV, and $\gamma_3 = 0.3$ eV are various hopping parameters defined in **Figure 9(b)**; and $<m,i;m',j>_0$, $<m,i;m',j>_1$, and $<m,i;m',j>_3$ run over pairs of sites within or between cells for the three types of hopping shown in **Figure 9 (b)**.

In the above Hamiltonians, on-site and hopping terms beyond the nanoribbon are set to be zero. Moreover, the effect of a transverse electric field is taken into account by including in the on-site energy term the linear-in-y atomic orbital energy shift due to the field.

Now consider the model system depicted in **Figure 9(c)** comprising left (first) and right (second) back-to-back filters. Let $n$ = number of atoms contained in each unit cell and $C_m$ = $n$-component tight-binding state vector in the $m^{th}$ unit cell. $C_m$ satisfies the following wave equation

$$(E - h_m)C_m - P^{(-)}C_{m-1} - P^{(+)}C_{m+1} = 0 \qquad (5)$$

where $E$ is the electron energy, $h_m$ the $n \times n$ matrix describing on-site orbital energy distribution as well as nearest-neighbor coupling within the $m^{th}$ unit cell, $P^{(+)}$ ($P^{(-)}$) the $n \times n$ matrix describing the nearest-neighbor coupling between $m^{th}$ and $(m+1)^{th}$ ($(m-1)^{th}$) unit cells. These various matrices are determined by the tight-binding Hamiltonian given earlier.

## 2. Formulation of the electron transmission problem

We formulate the transmission problem as follows. For VOI based valley valves, we are primarily interested in the transmission of electrons inside the pseudogap window. As shown earlier in **Figure 2(b)**, in this window there is only a single energy subband or, equivalently, for a given energy E, only a single state for each propagating direction with the Bloch form $exp(ik_x x)u_{kx}(x,y)$ where $u_{kx}(x,y)$ = cell-periodic function. This leads to a one-to-one correspondence between wave vectors and propagating directions, e.g., $k_x$ ($-k_x$) ↔ forward (backward) direction. We also note that $k_x$ above is a real number and the corresponding Bloch state has an extended probability distribution in the space.

For the transmission problem, we divide a VOI valve into three regions as shown in **Figure 9(c)**, which are defined by $m < 0$, $0 \leq m \leq N$, and $m > N$, where m denotes a unit cell index in the valve. The region with $0 \leq m \leq N$ contains the filter/filter interface while those with $m < 0$ or $m > N$ are some distance away from the interface. The wave function in the region with $0 \leq m \leq N$ generally includes both extended states (with $k_x$ being real-valued) and interface scattering-generated evanescent states (with $k_x$ being complex-valued) [36]. We take the regions with $m < 0$ or $m > N$ to be sufficiently far from the interface so that the evanescent states derived from the interface scattering already fade out before reaching the regions. Therefore, $C_{m<0}$ is given by only the extended state formed of superposition of incident and reflected Bloch states with real-valued wave vectors $k_x$ and $-k_x$, respectively. Similarly, $C_{m>N}$ is given by the transmitted Bloch state with real-valued wave vector $k_x$. Therefore,

$$C_{-1} = e^{-ik_x a_x} C_0^{(+)} + e^{ik_x a_x} C_0^{(-)},$$
$$C_{N+1} = e^{ik_x a_x} C_N, \qquad (6)$$

where $a_x$ is the lattice constant and $k_x$ the wave vector, $C_0 = C_0^{(+)} + C_0^{(-)}$, $C_N = C_N^{(+)}$. Superscripts (+) and (-) indicate forward and backward propagating components of the state, respectively. Specifically, $C_0^{(+)}$ represents the incident wave. Using the propagator formalism, Eqn. (5) can be transformed into one that describes the wave propagation from cell-0 to another cell:

$$C = GS. \tag{7}$$

Here, $G$ is Green's function with the inverse given by

$$G^{-1} = \begin{pmatrix} E-\tilde{h}_0 & -P^{(+)} & 0 & \cdots & 0 \\ -P^{(-)} & E-\tilde{h}_1 & -P^{(+)} & \cdots & \vdots \\ 0 & -P^{(-)} & E-\tilde{h}_2 & \ddots & 0 \\ \vdots & \vdots & \ddots & \ddots & -P^{(+)} \\ 0 & \ddots & 0 & -P^{(-)} & E-\tilde{h}_N \end{pmatrix} \tag{8}$$

where

$$\tilde{h}_m = \begin{cases} h_0 + P^{(-)}e^{ik_x a_x} & , \text{if } m=0; \\ h_N + P^{(+)}e^{ik_x a_x} & , \text{if } m=N; \\ h_m & , \text{if } 1 \leq m \leq N-1; \end{cases}$$

$$S = \left(P^{(-)}\left(e^{-ik_x a_x} - e^{ik_x a_x}\right)[C_0^{(+)}]^T, 0, \ldots, 0\right)^T;$$

$$C = \left([C_0]^T, [C_1]^T, \ldots, [C_N]^T\right)^T.$$

Eqn. (7) uses $C_0^{(+)}$ as the input and expresses the tight-binding state C in the structure in terms of $C_0^{(+)}$. As such, it provides a convenient framework for the calculation of transmission. Specifically, the $N^{th}$ component of Eqn. (7) reads

$$C_N = G_{N0} S_0. \tag{9}$$

With $S_0 \propto C_0^{(+)}$, Eqn. (9) relates the transmitted wave $C_N$ to the incident wave $C_0^{(+)}$.

$G_{N0}$ here is calculated efficiently using the recursive Green's function algorithm [49] as follows. In brief, instead of inverting the operator ($EI - H$) ($E$ = electron energy, $I$ = identity matrix, $H$ = tight-binding Hamiltonian) to solve for the whole set of propagator amplitudes, e.g., {$G_{mn}$'s, where $m$ and $n$ are arbitrary unit cell indices}, one calculates only two types of amplitudes, e.g. {$G_{m0}$, $G_{mm}$} for each unit cell ($m$ = cell index), using two recursive relations, with one expressing $G_{mm}$ in terms of $G_{m-1,m-1}$ and the other $G_{m0}$ in terms of $G_{m-1,0}$ and $G_{mm}$, in an iteratively fashion. The calculation starts from cell 0, proceeds to the next, and etc., until it reaches cell $N$ and finally obtains $G_{N0}$. As such, it constitutes an efficient algorithm for the calculation of transmission. For details, please see Reference [49].

Once $C_N$ is obtained, we compute the electron transmission from the ratio of output probability current to input probability current. This computation requires the expression of probability current operator in the tight-binding model similar to those used in discrete models [50–52]. Below we provide a brief discussion of the probability current operator used in this work.

By the probability conservation law, we write

$$\begin{aligned} \frac{\partial}{\partial t} P_m &= \frac{\partial}{\partial t} \sum_i C_{m,i}^* C_{m,i} \\ &= -\frac{1}{i\hbar}\left(C_{m-1}^+[P^{(+)}]C_m - C_m^+[P^{(-)}]C_{m-1}\right) \\ &\quad -\frac{1}{i\hbar}\left(C_{m+1}^+[P^{(-)}]C_m - C_m^+[P^{(+)}]C_{m+1}\right) \\ &= -[J^{(-)}(m) + J^{(+)}(m)] \end{aligned} \tag{10}$$

where $P_m$ is the probability of electron being located in unit cell $m$, $i$ denotes an atomic site in the unit cell,

$$J^{(-)}(m) = \frac{1}{i\hbar}\left(C_{m-1}^+[P^{(+)}]C_m - C_m^+[P^{(-)}]C_{m-1}\right) \tag{11}$$

representing the net probability current flowing through the interface between the $(m\text{-}1)^{th}$ and $m^{th}$ unit cells, and

$$J^{(+)}(m) = \frac{1}{i\hbar}\left(C_{m+1}^+[P^{(-)}]C_m - C_m^+[P^{(+)}]C_{m+1}\right) \tag{12}$$

representing the net probability current flowing through the interface between the $m^{th}$ and $(m+1)^{th}$ unit cells. The forgoing derivation has made use of the wave equation in (5) as well as the relation $[P^{(-)}]_{ji} = \left([P^{(-)}]_{ij}\right)^* = [P^{(+)}]_{ij}$, where $j$ is an atomic site index similar to $i$. We consider only stationary states, for which $J^{(-)}(m) + J^{(+)}(m) = 0$. From the above, we identify the probability current operator as

$$[j]_{m+1,m} = \frac{1}{i\hbar}\left([P^{(-)}]_{m+1,m} - [P^{(+)}]_{m,m+1}\right), \tag{13}$$

with the shorthand notation $[\ldots]_{a,b}$ denoting that the matrix inside the square bracket is to be multiplied with cell-$a$ state $C_a^+$ on the left and cell-$b$ state $C_a$ on the right. In our study since both the incident state and transmitted state are characterized by a single Bloch wave vector $k_x$, the above form can be written in terms of a new notation of current operator as

$$[\bar{j}]_{m,m} \equiv \frac{1}{i\hbar}\left([P^{(-)}]_{m,m}e^{-ik_x a_x} - [P^{(+)}]_{m,m}e^{ik_x a_x}\right) \tag{14}$$

with which the evaluation of current with a single cell state is possible.

Finally, in terms of the operator defined in Eqn. (14), we evaluate the electron transmission as

$$T(E) = C_N^+ \left[\overline{\overline{J}}\right] C_N \Big/ \left(C_0^{(+)}\right)^+ \left[\overline{\overline{J}}\right] C_0^{(+)}. \quad (15)$$

### 3. Numerical test of the formulation

A numerical test is presented below in regard to the required length of middle region ($0 \leq m \leq N$) for exclusion of evanescent states from regions $m < 0$ and $m > N$. In **Figure 10**, electron transmissions are calculated for two two-filter valves, e.g., SLG-I and BLG-I. Two filter lengths, namely, $L+L_s$ = 596 nm and 809 nm are considered for each valve. For either valve, as shown in the figure, transmission curves corresponding to the two filter lengths merge nicely, indicating that a good numerical convergence is already reached at the filter length of 596 nm or, equivalently, that evanescent states already vanish beyond this length.

### 1. Type-I and Type-II channels in SLG or BLG

**Figures 11-14** show electron transmissions in SLG-I, BLG-I, SLG-II, and BLG-II, respectively. In all cases we focus on channel states in the pseudogap window between the first and second conduction subbands, plot the corresponding electronic subband structure in and around the pseudogap, and calculate the electron transmission for electrons of the first subband with energy in the pseudogap window. The transmissions in P (parallel) configurations are always unity, which corresponds to the fundamental unit of quantum conductance, i.e., $2e^2/h$, in a typical Q1D quantum wire. On the other hand, in the cases presented, the transmissions in AP (anti-parallel) configurations can be as low as ~10%. Together they indicate the existence of a large on-off contrast in transmissions between P and AP configurations and confirm the presence of VOI-derived filtering effect.

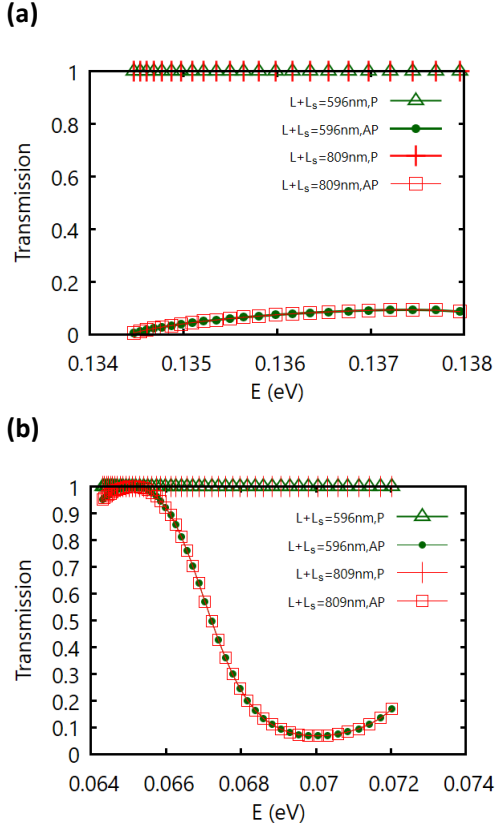

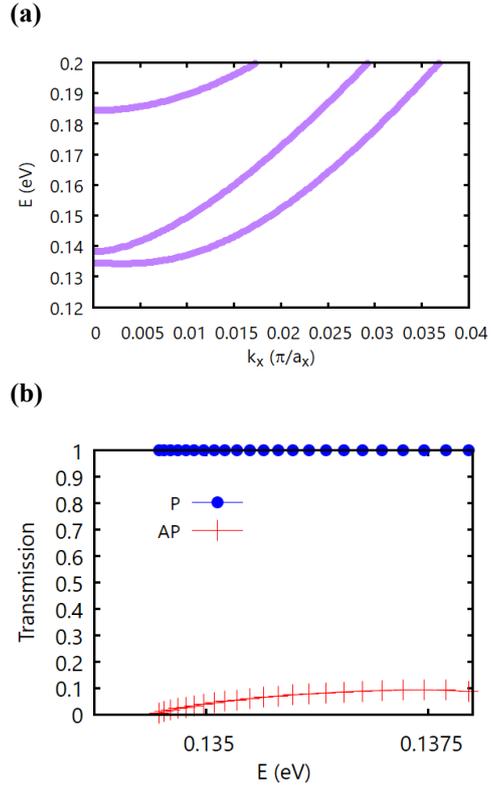

**Figure 10.** Convergence of transmissions in P and AP configurations, at two different filter lengths for **(a)** SLG-I and **(b)** BLG-I.

**Figure 11.** The case of SLG-I. **(a)** Conduction subbands. **(b)** The transmissions in P and AP configurations.

### IV. RESULTS

Below we present our numerical results of electron transmission in various cases.

We find that the on-off contrast is generally smaller in Type-II channels than in Type-I channels. This overall trend can be attributed to a typical reduction in valley filtering efficiency going from Type–I to Type–II channels, due to generally reduced band gap fluctuations with staggered band alignment in comparison to those with straddling band alignment. As shown in **Figure 13**, the contrast decreases significantly in the specific case of SLG-II with barrier gap ~ channel gap, where the barrier gap is elevated only by 0.02 eV relative to the channel gap. The insignificant contrast can be argued for based on approximate Ehrenfest's theorem in the limit where the band gap fluctuation vanishes. Within the effective one-band model of gapped SLG[6], the Hamiltonian in this limit is given by $H \approx v_F^2 p^2/2\Delta + V$, and the theorem takes the form $\langle \partial p_y / \partial t \rangle \approx -\langle \nabla_y V \rangle$. When the forgoing approximate theorem is applied to a channel-confined state in the uniform gap limit, it results in a nearly vanishing quantum average $\langle \nabla_y V \rangle \approx 0$ implying weak Rashba valley splitting or, equivalently, diminished valley contrast in constituent filters and the valve. On the other hand, a BLG structure or a finite gap variation in the SLG structure invalidate the above filtering, as reflected in the sizable on-off contrast in **Figures 11**, **12** and **14** for SLG-I, BLG-I and BLG-II.

It is also interesting to compare SLG and BLG based valves, for transmissions in the AP configurations. In **Figures 11** and **13**, we find that transmissions in SLG valves exhibit a slow rise with increasing energy near the on-set and a global suppression across the entire pseudogap window. On the other hand, in **Figures 12** and **14**, transmissions in BLG valves show steep rise near the on-set and notable suppression near the upper end of pseudogap window. The feature of low transmission near the upper pseudogap edge is common to both SLG and BLG valves. It is actually a universal behavior consistent with the description of VOI valley filtering given earlier, namely that the electron state becomes increasingly valley polarized away from the lower pseudogap edge. However, details of the variation near transmission on-set are layer number-dependent and determined by two competing factors as follows. First, the electron state near lower pseudogap edge is 50:50 valley mixed and this strong mixing invalidates the simple, total on-total off valve picture that

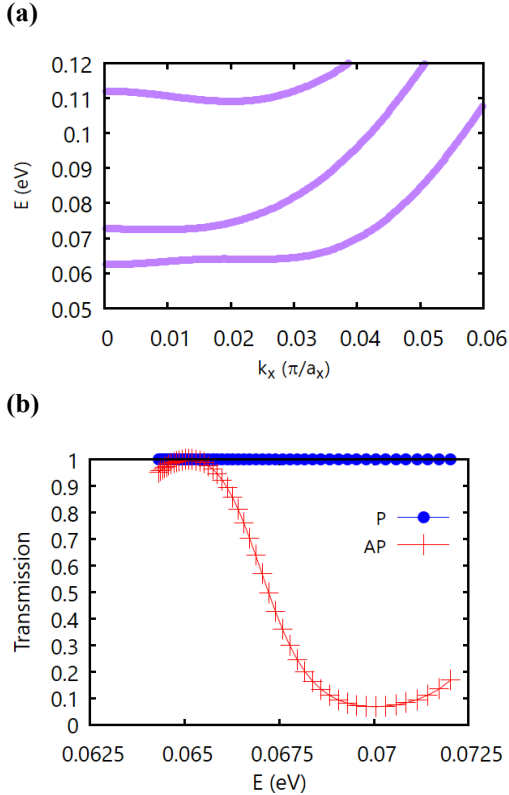

**Figure 12.** The case of BLG-I. **(a)** Conduction subbands. **(b)** The transmissions in P and AP configurations.

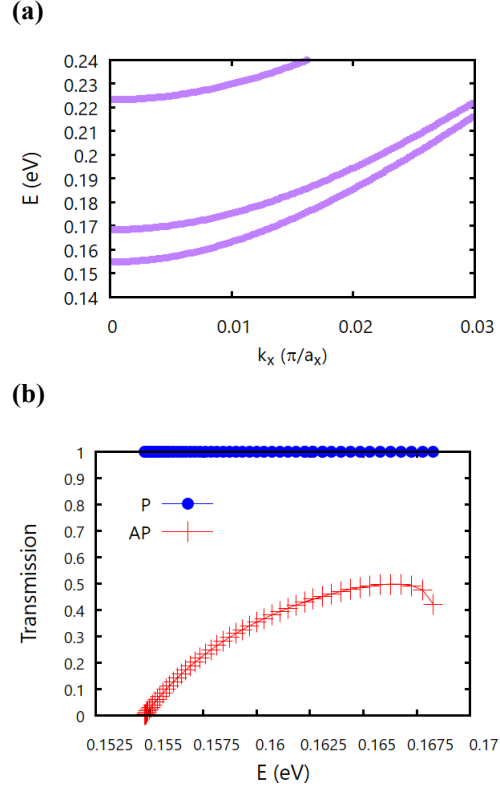

**Figure 13.** The case of SLG-II. **(a)** Conduction subbands. **(b)** The transmissions in P and AP configurations.

argument and thus permit notable valley

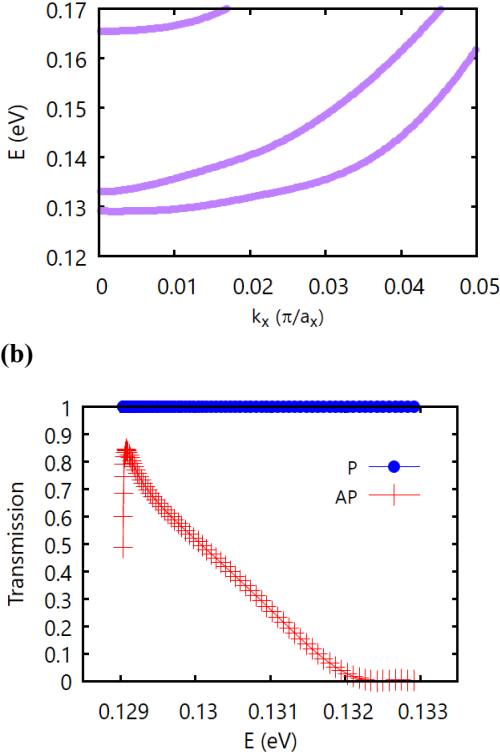

**Figure 14.** The case of BLG-II. **(a)** Conduction subbands. **(b)** The transmissions in P and AP configurations.

assumes an ideal, 100% filtering efficiency in constituent filters. Breakdown of the picture implies a sizable transmission near the on-set. Second, however, based on another picture one would expect the transmission to decrease with decreasing energy, as argued below from the perspective of electron transmission as a continuous function of electron energy. Consider the variation of transmission with decreasing energy, for a state in the pesudogap. This state lies in the first conduction subband but, with decreasing energy, it eventually exits the subband and enters the valence-conduction band gap below the subband, where the transmission is totally forbidden. By interpolation between a subband state and a gap state, one thus expects a decrease in the transmission with decreasing energy, at least when the energy lies sufficiently close to the on-set. Generally, details of the competition between two forgoing factors vary with the bulk band structure. For SLG structures, as shown in **Figures 11** and **13**, the second factor dominates resulting in a monotonously increasing transmission with increasing energy. For BLG structures, the first factor is dominant for energy slightly above the on-set, leading to an almost monotonously decreasing transmission with increasing energy, as shown in **Figures 12** and **14**.

### 2. Correspondence between filtering efficiency and switching capability

The off-switching capability of a valley valve primarily derives from the correlation between low electron transmission and valley polarity disparity between constituent filters. When the valve is placed in the AP configuration, the disparity develops with a degree depending on the filtering efficiency of filters. In **Figure 15**, we show both corresponding valley polarizations in constituent filters and electron transmissions in the AP configuration, for SLG-I, BLG-I, SLG-II, and BLG-II, with polarizations calculated by the subband structure theory developed in Reference [36] and employed in **W-I** while transmissions by the theory of present work. In each case, we see that pseudogap windows obtained with the two different methods roughly agree in size and range. Moreover, the concurrence of weak (strong) polarization with high (low) transmission is evident when reading **Figures 15(a)** and **15(c)** for SLG based valves. Specifically, in **Figure 15(c)** where the structure features a weak band gap variation, we see the coincidence of small polarization ~ 10% with large transmission ~ 50%. Similarly, in **Figures 15(b)** and **15(d)** for BLG based valves, we see that comparable valley polarizations result in comparable transmissions, again confirming the polarization-transmission correlation. We note, however, that there is a slight deviation from the correlation - while the valley polarization near high-energy end of pseudogap window is slightly larger in **Figure 15(b)** than in **Figure 15(d)**, the transmission is actually slightly lower in **Figure 15(d)** than in **Figure 15(b)**, showing a secondary factor at work in controlling the transmission in AP configuration. For valves here that are formed of back-to-back filters, this factor comes from additional wave function mismatch apart from the valley polarity disparity between two filters. Due to antiparallel fields in the configuration, envelops of subband state wave functions tilt towards opposite sides in the two filters, with details of resultant distributions depending on transverse confinement potential profiles.

**(a)**

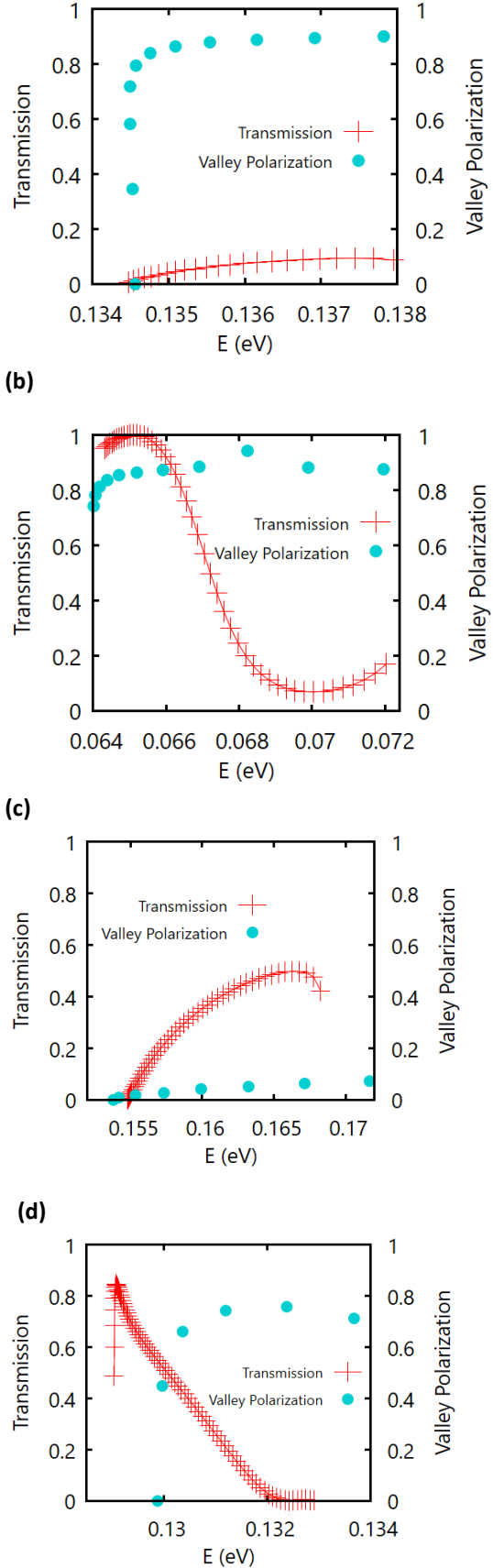

**Figure 15.** Correspondence between polarizations and transmissions in the AP configuration, in **(a)** SLG-I, **(b)** BLG-I, **(c)** SLG-II, and **(d)** BLG-II.

Therefore, an envelope function disparity develops between the filters and, along with valley polarity disparity, gives rise to the total wave function mismatch that determines the amount of electron backscattering at the filter-filter interface. In the cases of **Figures 15(b)** and **15(d)**, with comparable polarizations, the valley polarity disparity is about the same. On the other hand, since the two cases involve different channel types and transverse potential energy drops, e.g., 0.06 eV and 0.08 eV, respectively, the strength of envelope function disparity varies between the two cases ending up in tipping the scale and resulting in a weak breakdown of the polarization-transmission correlation.

## 3. Effect of transverse fields

The effect of transverse electric fields is presented in **Figure 16**. With the transmissions in P configuration being always unity as noted in **Figures 11-14**, below we investigate SLG and BLG valves with Type-I channels in the AP configuration. We note two points. First, the pseudogap window is expected to shift towards low energy with increasing electric field. This is attributed to electric field-induced energy lowering of the first conduction subband, which derives from level repulsion between the first and second subbands in the presence of a perturbation, e.g., the electric field. Second, based on the VOI filtering principle explained earlier, Rashba valley splitting is expected to increase with increasing electric field. As a result, increasing electric field would raise valley polarization or, equivalently, reduce overall electron transmission according to the polarization-transmission correlation. **Figure 16** bears out such an expectation. In particular, **Figure 16(a)** shows clearly a decrease in transmission with increasing electric field. On the other hand, there are some differences between SLG and BLG valves as shown in **Figures 16(a)** and **16(b)**. As noted earlier in the discussion of **Figures 12** and **14**, with BLG valves in the AP configuration there is a steep rise in transmission near the on-set. In **Fig 16(b)**, this steep rise at the on-set cushions the fall in transmission with increasing field. However, with increasing field, an expansion in the energy range of low transmission is still evident in the high-energy tail. Overall, **Figure 16** demonstrates the increased electrical controllability of VOI based valley valves with increasing field strength.

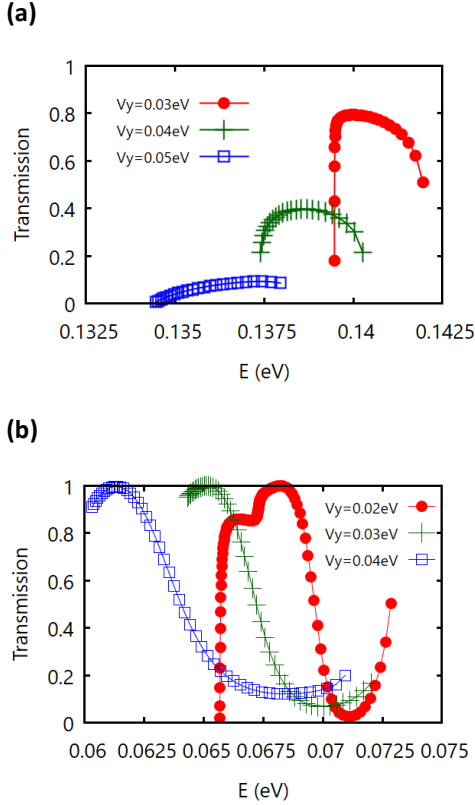

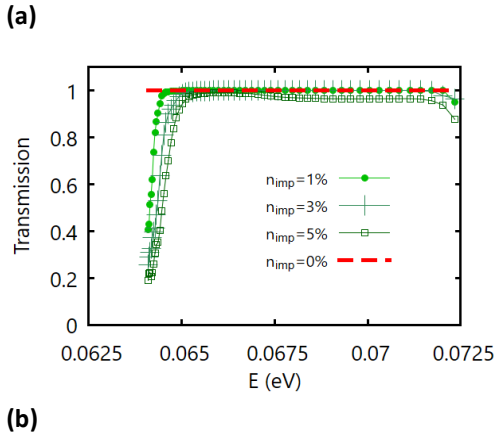

**Figure 16**. VOI filtering capacity in different field strengths, in **(a)** SLG-I with $V_y$ = 0.03, 0.04, and 0.05 eV, and in **(b)** BLG-I with $V_y$ = 0.02, 0.03, and 0.04 eV. Note that the transmissions at $V_y$ = 0 eV are the same as those in the P configuration and given by unity (not shown).

## 4. Impurity scattering

We present the impurity scattering effect for the valley valve BLG-I. **Figures 17(a)** and **17(b)** show, respectively, transmissions in P and AP configuration with long-range impurity scattering, while **Figures 18(a)** and **18(b)** show those with short-range impurity scattering. The result is averaged over 100 (200) configurations in the case of long (short) – range impurities.

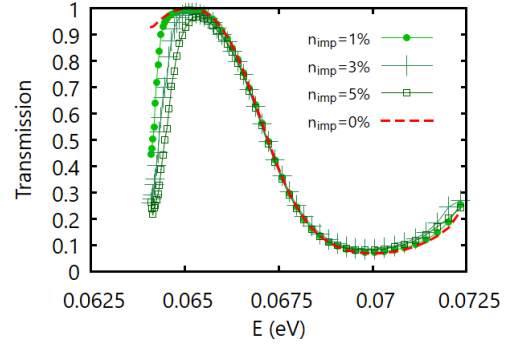

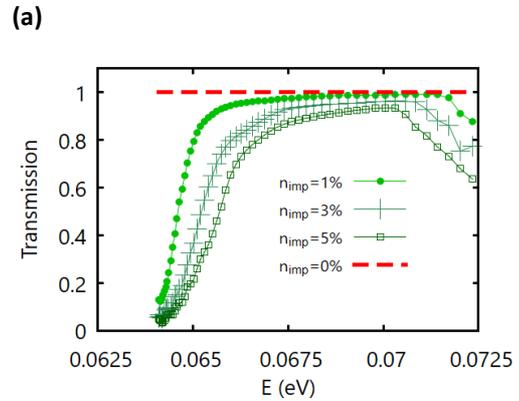

**Figure 17.** Long-range impurity scattering in BLG-I. **(a)** Transmissions in the P configuration. **(b)** Transmissions in the AP configuration. Impurity densities ($n_{imp}$) are taken to be 0%, 1%, 3%, and 5%, respectively.

Due to the suppression of impurity scattering by valley-wave vector locking in the filters, robust valve switching effect is expected in the presence of limited impurity scattering. In order to test the strength of robustness, four levels of impurity density, namely, 0%, 1%, 3%, and 5% are investigated.

Generally, we find that with increasing impurity concentration the transmission increasingly deviates from that of the clean structure. However, we see that a considerable on-off contrast in the transmission sustains in all cases considered here. Note that the contrast is relatively reduced in **Figure 18** when compared to that in **Figure 17**, due to relatively frequent intervalley scattering by short-range impurities.

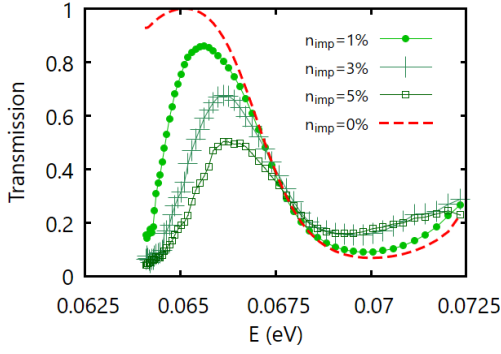

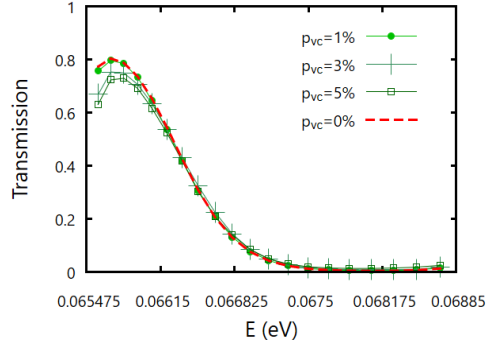

**Figure 18.** Short-range impurity scattering in BLG-I. **(a)** Transmissions in the P configuration. **(b)** Transmissions in the AP configuration. Impurity densities ($n_{imp}$) are taken to be 0%, 1%, 3%, and 5%, respectively.

**Figure 19.** Edge roughness scattering in BLG-I valve. **(a)** Transmissions in the P configuration. **(b)** Transmissions in the AP configuration. $p_{vc}$ (probability of vacancy cluster occurrence) is taken to be 0%, 1%, 3%, and 5%.

Furthermore, in a typical case, the contrast is slightly more robust at the high-energy tail than near the on-set of the transmission curve. This is expected since the corresponding electron state is relatively valley mixed at the on-set and relatively valley purified at the tail, resulting in a more effective valley-wave vector locking-induced suppression of impurity scattering at the tail than at the on-set.

## 5. Edge roughness

Now we present the effect of edge roughness for the valley valve BLG-I. **Figures 19(a)** and **20(a)** show transmissions in the P configuration and **Figures 19(b)** and **20(b)** show those in the AP configuration. Each result is averaged over 100 configurations.

Due to valley-wave vector locking, the valve is expected to show robustness against limited edge roughness scattering. In order to test the strength of robustness, four levels of edge roughness are considered in **Figure 19**, with $p_{vc}$ (probability of a vacancy cluster occurring) being given by 0, 1%, 3%, and 5%. As shown in the figure, edge roughness introduces a limited deviation in transmission from that with $p_{vc} = 0$, with the deviation increasing with increasing $p_{vc}$.

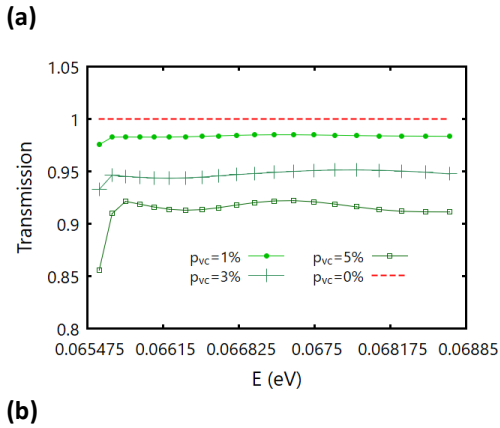

In **Figure 20**, we compare two BLG-I valves with the same w2 (channel width) at 7.5 nm, the same $p_{vc}$ at 1% but different w (total ribbon width) at 17.3nm and 24.7nm, respectively, which correspond to different edge-channel distances. We see that the effect is stronger in the narrow ribbon structure than in the wide ribbon structure, which is expected since passing electrons collide with boundaries more often in the narrow ribbon than in the wide ribbon. This quantitative result here provides useful clues to the design of VOI valley valves as well as general nanoribbon-based structures.

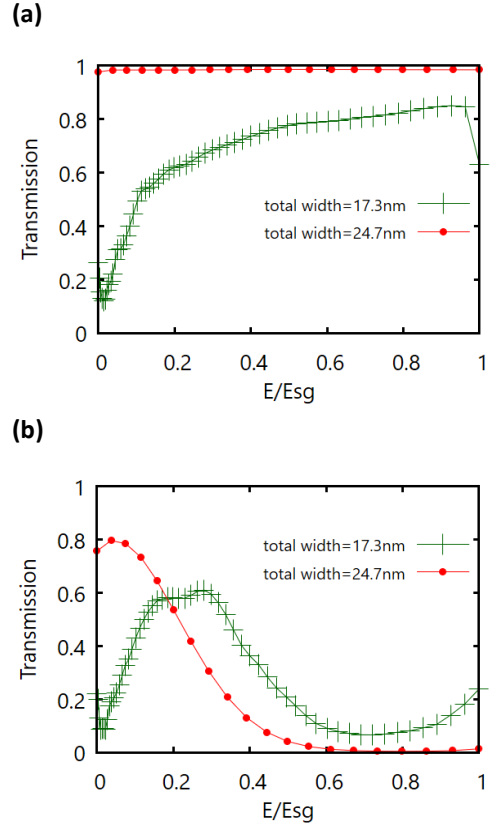

**Figure 20.** Comparison about the effect of edge roughness between two BLG-I valves with the same channel width at 7.50 nm, the same $p_{vc}$ at 1% but different total ribbon widths at w = 17.3nm and w = 24.7nm, respectively. **(a)** Transmissions in the P configuration. **(b)** Transmissions in the AP configuration. For comparison, we normalize the size ($E_{sg}$) and range of pseudogap window to unity and (0,1), respectively, and use the relative electron energy "$E/E_{sg}$" in the plot. $E_{sg}$ = 8.24 (3.12) meV for w = 17.3 (24.7) nm.

**6. Three-filter valves**

We present transmissions in three-filter valves. In the P configuration, all valves manifest full transparency for electron transmission. However, as shown in **Figure 21**, in the AP configuration, the middle filter serves as a cavity and transmission peaks emerge, due to electron wave resonances in the cavity.

Specifically, **Figure 21(a) ((b))** presents transmissions in the AP configuration for both BLG-I (BLG-II) and a corresponding three-filter valve, e.g., BLG-I (BLG-II) inserted with a middle filter 42.6nm (51.1nm) long. **Figure 21(c) ((d))** re-plots the three-filter transmission in **Figure 21(a) ((b))** but with the horizontal variable "E" (electron energy) converted to the corresponding wavelength, e.g., "$2\pi/k_x$" ($k_x$ = electron wave vector). As shown in **Figures 21(c) and 21(d)**, the peak labelled "Resonance" occur at a specific wavelength that correlates to the middle filter length by a simple ratio. Such a correlation applies to other resonances as well, when the increasing red shift of resonance with increasing electron energy (or decreasing wave length) due to limited barrier height is accounted for. Overall, the above correlation supports the identification of these peaks as Fano-Fabry-Perot resonances in association with electron transmission via quasi-bound states (QBS) in the middle filter.

Transmission resonances in three-filter valves affect the on-off transmission contrast as follows. Take **Figure 21(a)** as an example. A peak arises in the energy range between 0.068 eV and 0.069 eV. In association with it, a reduction of transmission next to the peak occurs when compared to the two-filter transmission, due to the presence of double valley polarity disparity in the three-filter valve as opposed to that of only single disparity in the two-filter valve. As a result, the on-off contrast in transmission above 0.069 eV is enhanced in the three-filter valve over that of the two-filter valve. A similar observation applies to **Figure 21(b)** as well when comparing the two- and three- filter transmissions.

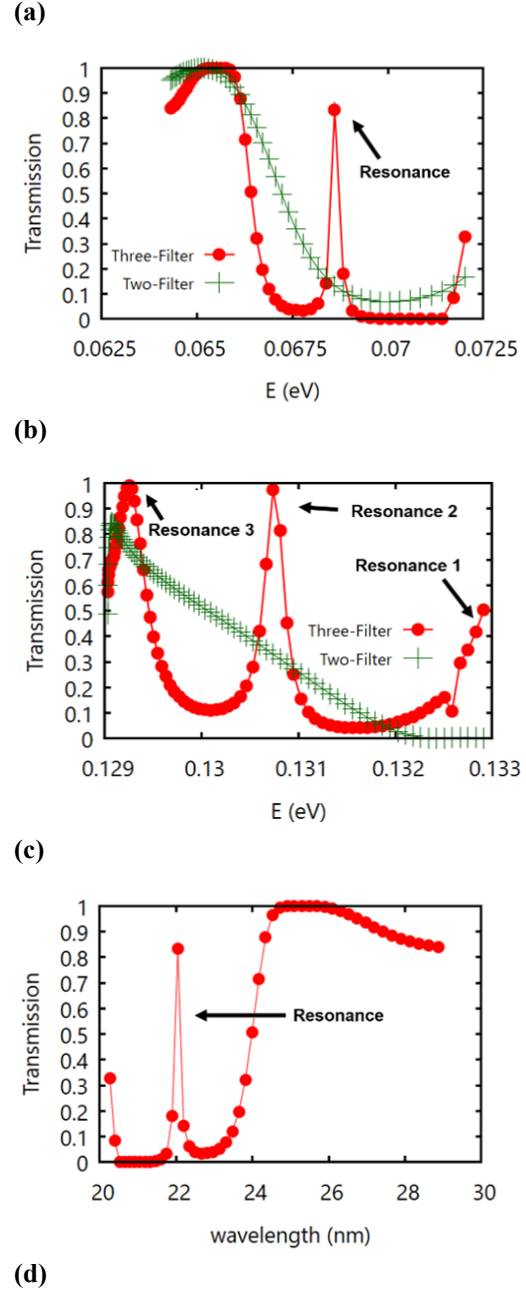

(a)

(b)

(c)

(d)

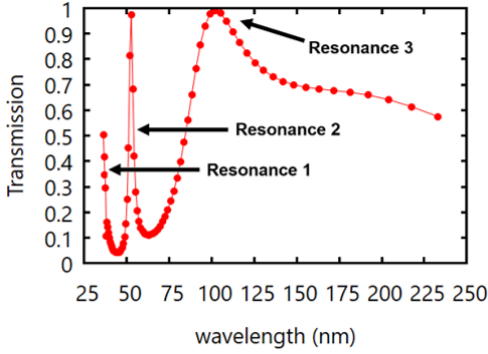

**Figure 21.** Transmissions of three-filter valves in the AP configuration, for **(a)** BLG-I inserted with a middle filter 42.6nm long, and **(b)** BLG-II inserted with a middle filter 51.1 nm long. For comparison, transmissions in corresponding two-filter valves are also shown in **(a)** and **(b)**. **(c)** Re-plot of the three-filter transmission in **(a)** but with the variable E (electron energy) converted to the corresponding wavelength. **(d)** Re-plot of the three-filter transmission in **(b)** but with the variable E converted to the corresponding wavelength.

Overall, with three-filter valves, the emergence of resonances modifies the transmission and such modification could bring benefits to on-off switching. In the case of **Figure 21(a)**, one could, for example, utilize the energy range above 0.069 eV for on-off switching in order to benefit from the enhanced on-off transmission contrast in that energy range.

## 7. Inter-filter spacer effect

Similar to the middle filter in a three-filter structure, inter-filter space in a valve also plays the role of a cavity that can lead to Fano-Fabry-Perot resonances in the transmission. For VOI-based valves, additional effects could be caused by the presence of inter-filter region. For example, if the region consists of a nanoribbon, then valley polarized electrons exiting the first filter would, when passing through the spacer, suffer from inter-valley mixing due to boundary scattering in the ribbon. As a result, they would be depolarized before going into the next filter. When coupled with Fano-Fabry-Perot resonances, it could considerably affect transmissions in both P and AP configurations.

In **Figures 22**, we study the effect of an inter-filter spacer on electron transmission, with respect to a variation in the spacer width and length. Two similar structures are considered and compared. The first one is labelled as S22-A and derived from BLG-I with a nanoribbon spacer 17.3nm wide and 21.3nm long inserted between the two filters. The second structure, labelled as S22-B, is obtained from the first one by increasing the nanoribbon width to 24.6nm. **Figures 22(a)** and **22(b)** show energy subbands in both the filter and the spacer ribbon in the two cases. We see that since both spacers are wider than the filter's channel, subbands in both spacers shift downwards relative to those in the filter, as the result of weaker confinement in the spacers, with the average shift being larger in **Figure 22(b)** with S22-B than that in **Figure 22(a)** with S22-A. **Figures 22(c)** and **22(d)** present corresponding transmissions for the two structures, respectively. In both cases clear signatures of Fano-Fabry-Perot resonances are demonstrated in association with QBS in the inter-filter space, for transmissions in both P and AP configurations. However, due to different spacer ribbon widths, the strength of ribbon edge scattering induced-intervalley mixing varies between the two cases and affects the relative order of transmissions in P and AP configurations in the two cases. While typically the transmission for P configuration is larger than that for AP configuration, this relative order is disrupted with S22-A. The disruption can be attributed to the valley depolarization due to strong inter-valley mixing in the ribbon spacer in S22-A, which invalidates the simple on-off valve switching picture based on ideal valley filtering. On the other hand, for S22-B, as shown in **Figure 22(d)**, with a wide ribbon spacer it tends to maintain the usual relative order of

**(a)**

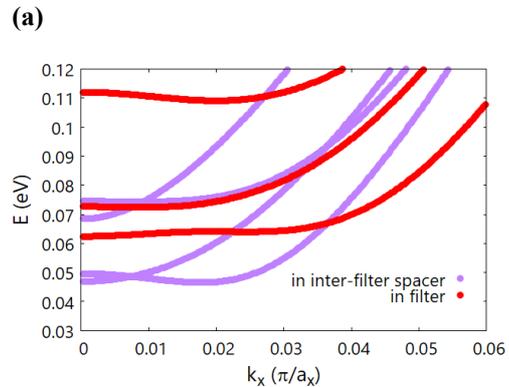

**(b)**

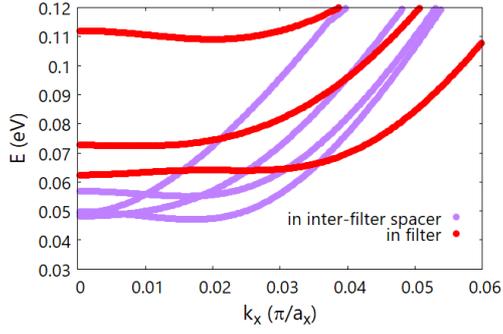

(c)

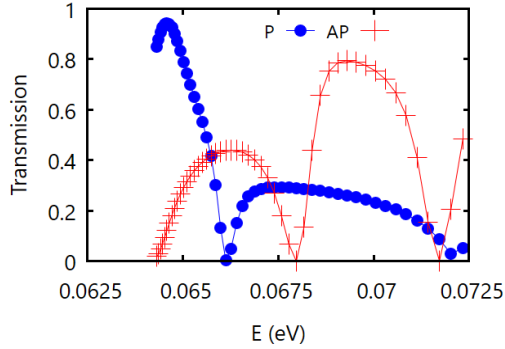

(d)

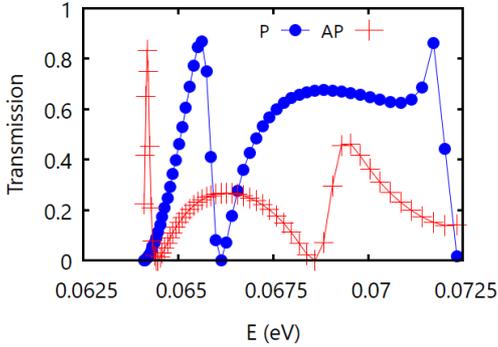

**Figure 22.** Inter-filter spacer effect. **(a)** Conduction subbands in the filter and spacer of S22-A. **(b)** Conduction subbands in the filter and spacer of S22-B. **(c)** Transmissions in P and AP configurations for S22-A. **(d)** Transmissions in P and AP configurations for S22-B.

transmissions because of relatively weak inter-valley mixing or depolarization in a wide spacer.

## 8. Long distance valley transport

In **Figure 23**, we discuss the issue of long distance valley transport between two filters. Two structures, namely, S23-A and S23-B, are investigated here. S23-A is derived from BLG-I, with an inter-filter spacer inserted that is formed with confinement barriers to provide a Q1D channel for electron passage. The spacer channel is 25.56 nm long and specifically chosen to be as wide as the filter channel, but with thinner barriers that are 3.07 nm wide. S23-B is obtained from S23-A by increasing the spacer channel length by about a factor of ten to 256 nm. With the above choice of channel width and barrier width for the spacer, the low-lying spacer subbands in both structures are roughly aligned with corresponding filter subbands, as shown in **Figure 23(a)**. Such subband matching significantly facilitates the valley transport between two filters. In the case of S23-A, as shown in **Figure 23(b)**, the transmissions in P and AP configurations are able to follow closely, in spite of the presence of Fano-Fabry-Perot resonances, those of BLG-I without any spacer. In the case of S23-B with an extremely long spacer, as shown in **Figure 23(c)**, the number of Fano-Fabry-Perot resonances increases rapidly, but the correlation in transmissions between S23-B and BLG-I is manifest, indicating that the valley polarized current exiting the first filter is able to last over a long distance. In contrast to what we have seen in **Figure 22**, where a strong valley depolarization is present in S22-A with a nanoribbon spacer, the presence of barriers in

(a)

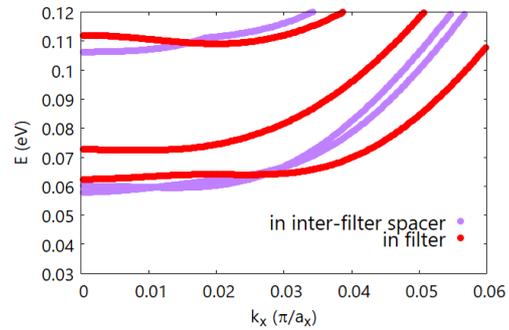

(b)

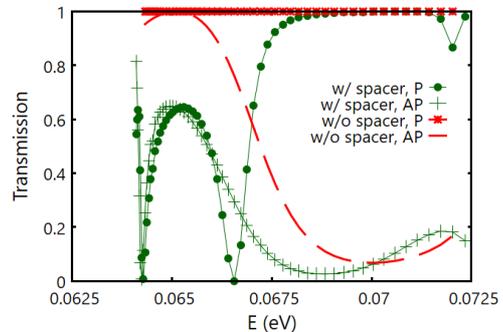

(c)

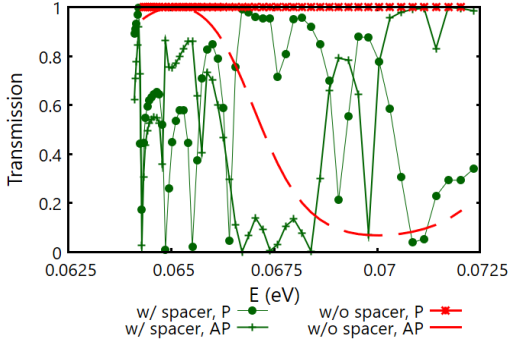

**Figure 23.** Long distance valley transport. **(a)** Alignment of filter and spacer subbands. **(b)** The transmissions with and without the presence of spacer in S23-A. **(c)** The transmissions with and without the presence of spacer in S23-B.

the spacer of S23-B now effectively protects passing electrons from ribbon edge scattering in the spacer thus allowing for the valley current here to sustain over a long distance.

## 9. Uniform strain

In the presence of a uniaxial strain along the AGNR, a straightforward analysis of graphene band structure using the tight-binding model [53] shows that original Dirac valleys at K and K′ are shifted to K + $\delta k$ and K′ - $\delta k$, respectively, with the shift $\delta k$ roughly scaling with the strain or $(r-1)$, where $r = t'/t$ being the ratio of zigzag hopping to armchair hopping. Such a shift is reminiscent of wave vector shift induced by the vector potential of a magnetic field and can modulate the valley transport.

In passing, we note that the hopping ratio $r$ is introduced here to primarily serve as a convenient strain parameter, which describe the strain in such a way that it can directly be included in the tight-binding calculation. To establish the connection between $r$ and the actual strain, we consider the example of a 1D monatomic chain. In the example, if one models the nearest neighbor hopping by a power law and takes $t' = t\,(a/a_0)^{-n}$ ($t\,(a_0)$ = hopping (lattice constant) of the unstrained chain, $t'\,(a)$ = hopping (lattice constant) of the strained chain, $n$ = some positive number), then in the weak strain limit, $r-1$ conforms with the actual strain in the sense that $r-1 \approx -n\,(a-a_0)/a_0$ in the limit, essentially the same as the actual strain given by "$(a-a_0)/a_0$" except for an overall pre-factor.

**Figure 24** presents the effect in SLG-I in the AP configuration. While the transmission in P configuration remains at unity, that in the AP configuration manifests a quasi-periodic oscillation as shown in the figure, when the strain is varied by taking r = 1.00, 1.01, 1.02, 1.03, 1.04, 1.05, 1.06, 1.07, and 1.08. This oscillation is explained below.

The shift $\delta k$ has the following effect on electron states. When considering the transverse quantization of electron states in an AGNR, it leads to the emergence of a corresponding phase factor $e^{i\phi}$, with $\phi = w \cdot \delta k/2$ (w = ribbon width), in the condition for transverse quantization of electron states. Therefore, variation of the strain (or $\delta k$) results in an interesting quasi-periodic modulation of valley transport somewhat analogous to the Shubnikov de Haas oscillation, as shown in **Figure 24**. This modulation could provide potential applications in NEMS (nanoelectromechanical systems).

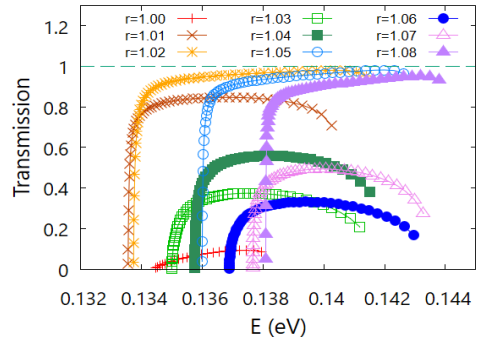

**Figure 24.** Transmissions in AP configuration for SLG-I under various uniaxial strains corresponding to r = 1.00, 1.01, 1.02, 1.03, 1.04, 1.05, 1.06, 1.07, 1.08, with r = $t'/t$ being the ratio of zigzag hopping to armchair hopping.

## V. SUMMARY

In this work we have studied the valley-dependent transport of electrons in valley valves within the tight-binding model of graphene, using VOI based valves as examples. Recursive Green's function algorithm has been applied to the calculation of electron transmission coefficients.

General features of valley transport in valves have been demonstrated. In the case of a two-filter valve, it confirms the existence of a pronounced on-off contrast in transmission between the two valve configurations, namely, one with identical and the other with opposite valley polarities in the two constituent filters. Three-filter valves are compared to two-filter ones, with the result showing a better on-off contrast for three-filter valves. Fano-Fabry-Perot type resonances in the transmission are illustrated, in the cases of three-filter valves or two-filter ones with inter-filter space, which indicates the necessity of a careful structural design of valves for a good on-off switching performance. Valves in the presence of either long- or short- range impurity scattering potentials have been investigated. It shows robustness against limited degree of impurity scattering due to the valley-wave vector locking in constituent valley filters.

Features specific to VOI based valves have also been illustrated. For example, it has been shown that the electron transmission in AP configuration can be tuned through the gate bias difference transverse to the channel, therefore allowing for electrical control of on-off switching with VOI based valves. Moreover, the on-off contrast is shown to exist in both SLG and BLG based valves with either Type- I or II channels. Generally, Type-I channels exhibit better contrast due to its relatively large band gap variation in comparison to that in Type-II channels. Edge roughness scattering effects have been investigated, too, which demonstrates tolerance of VOI valves with respect to limited edge roughness. In valves with nanoribbon inter-filter spacers, we have shown that a design with wide ribbon spacers can reduce the boundary-caused valley depolarization of electrons in spacers. Last, uniaxial strain effects have been examined, showing significant modulation of the electron transmission with a quasi-periodic strain dependence.

In conclusion, rich, interesting physics has been demonstrated for valley transport in valley valves. In the case of VOI-based valves, robust, pronounced on-off contrasts in electron transmission have been illustrated, implying a viable path offered by such valves to the implementation of electro-valleytronic circuits.

### ACKNOWLEDGMENT

We would like to thank Pei-Shiuan Wu, Tzu-Chi Hsieh, Hui-Po Huang, and Yen-Chun Chen for discussions. We acknowledge Prof. Mei-Yin Chou for providing computational resources, and support of MoST, ROC through Contract No. MOST 109-2811-M-007-561.